\documentclass[12pt,a4paper]{article}
\usepackage{amsmath}
\usepackage{amsfonts}
\usepackage{amssymb}
\usepackage{textcomp}
\usepackage[dvips]{graphicx}
\usepackage{epsfig}
\usepackage{bm}
\usepackage{dcolumn}
\usepackage{amsfonts}
\usepackage{color}
\usepackage[left=2cm,right=2cm,top=2cm,bottom=2cm]{geometry}
\usepackage[left=2cm,right=2cm,top=2cm,bottom=2cm]{geometry}
\usepackage[tableposition=top]{caption}
\usepackage{subcaption}
\DeclareCaptionLabelFormat{gostfigure}{Fig. #2}
\DeclareCaptionLabelSeparator{gost}{.~~}
\begin{document}
\title{\textbf{Asymptotic cosmological regimes in scalar-torsion gravity with a perfect fluid}}
\author{Maria A. Skugoreva$^{1}$\footnote{masha-sk@mail.ru}, Alexey V. Toporensky$^{1,2}$\footnote{atopor@rambler.ru}\vspace*{3mm} \\
\small $^{1}$Kazan Federal University, Kremlevskaya 18, Kazan 420008, Russia\\
\small $^{2}$Sternberg Astronomical Institute, Lomonosov Moscow State University,\\
\small Moscow, 119991, Russia}

\date{ \ }
\maketitle

\begin{abstract}
We consider the cosmological dynamics of a nonminimally coupled scalar field in scalar-torsion gravity in the presence of hydrodynamical matter. The potential of the scalar field have been chosen as power law with negative index, this type of potentials is usually used in quintessence scenarios. We identify several asymptotic regimes, including de Sitter, kinetic dominance, kinetic tracker, and tracker solutions and study the conditions for their existence and stability. We show that for each combination of coupling constant and potential power index one of the regimes studied in the present paper is stable to the future.  
\end{abstract}

\section{Introduction}
~~~~Recently a new class of theories modifying General Relativity (GR) has started to attract great attention mainly in order to explain observation data indicating the accelerated expansion of the late Universe \cite{observ}. It is based on the rather old theory attributed to Einstein himself \cite{Einstein}, however, forgotten for decades since it is in fact not a separate theory from the viewpoint of equation of motion, but rather a reformulation of GR. Instead of torsion-free Levi-Civita connections it uses curvature free Weitzenb\"{o}ck connections \cite{Weitzenbock}. It is well known that zero-curvature connections allow for the existence of a path-independent definition of vector parallel transport, so this theory has got the name of the Teleparallel Equivalent of General Relativity (TEGR) \cite{TEGR,Hayashi:1979qx,JGPereira,Maluf:2013gaa}. What, however, has become clear rather recently is that the well-known modifications of GR, such as a theory of a scalar field nonminimally coupled to gravity, when constructed on the base of TEGR, lead to different equations of motion (for the reason of this see \cite{Bahamonde}). Such a theory has been intensively investigated in many papers during several recent years.

Generalizations of TEGR are usually constructed in the same way as modifications of GR. It is possible to generalize the Lagrangian replacing in it the torsion scalar $T$ with a function $f(T)$ \cite{Ferraro:2006jd,Linder:2010py}. Different types of cosmological scenarios appear in scalar-torsion gravity, namely, in the class of models with nonminimal coupling between the torsion scalar and the scalar field $\phi$ of the form $\xi T F(\phi)$, where $\xi$ is a coupling constant, $F(\phi)$ is some function of a scalar field \cite{Geng:2011aj,Xu:2012jf,Otalora:2013tba,Geng:2013uga}. Recently some other modifications of TEGR including a nonminimal derivative coupling to torsion \cite{Kofinas} and analogs of the Gauss-Bonnet invariant have appeared \cite{Kofinas:2014owa}. There are also modifications of TEGR with no direct analogs of GR modification \cite{Bahamonde}.

In our recent paper \cite{we} we applied dynamical system methods to scaler-torsion theory in order to find some cosmological asymptotic regimes and describe the corresponding phase portraits, focusing mainly on growing scalar field potentials.
The main qualitative result of that paper is that scalar-torsion coupling leads to much less variety of possible dynamical regimes than scalar-curvature coupling \cite{wesami,wevernov}. In the present paper we give an heuristic explanation of this feature, as well as show that more possibilities appears for decreasing scalar fields potentials, usually used in quintessence models. As such models have been created for a description of the late Universe, when the usual matter is important as well, we add hydrodynamical matter to the scalar field. We will use units with $\hbar=c=1$.

The paper is organized as follows: in Sect. 2 we present equation of motion in initial and expansion normalized variables, in Sect. 3 the analysis of stationary points and corresponding asymptotic solutions is given, and Sect. 4 gives a brief summary of results obtained. 

\section{Main equations}
~~~~We start with describing basic objects of teleparallel gravity.
In this theory the dynamical variables are the
tetrad fields (also called the vierbein fields) ${\mathbf{e}_A(x^\mu)}$; here Greek indices relate to space-time and capital Latin indices belong to the tangent space-time. The metric tensor is expressed in terms of  the tetrad as 
\begin{equation}  
\label{gmn}
g_{\mu\nu}=\eta_{\mathrm{AB}}\, e^A_\mu \, e^B_\nu,
\end{equation}
where $\eta_{\mathrm{AB}}=\mathrm {diag} (1,-1,-1,-1)$. For the definition of parallel transport of a vector the Weitzenb\"{o}ck connection \cite{Weitzenbock} is used, 
\begin{equation}
\label{GW}
\overset{\mathbf{w}}\Gamma^{\lambda}_{\mu\nu}\equiv e^{\lambda}_A\partial_{\mu}e^{A}_{\nu}.
\end{equation} 
Then the torsion tensor and the torsion scalar are given by
\begin{equation}
\label{Tlmn}
T^{\lambda}_{\mu\nu}\equiv\overset{\mathbf{w}}\Gamma^{\lambda}_{\nu\mu}-\overset{\mathbf{w}}\Gamma^{\lambda}_{\mu\nu}=e^{\lambda}_A(\partial_\mu e^A_{\nu}-\partial_\nu e^A_{\mu}),
\end{equation}
\begin{equation}
\label{T}
T\equiv\frac{1}{4}T^{\rho\mu\nu}T_{\rho\mu\nu}+\frac{1}{2}T^{\rho\mu\nu}T_{\nu\mu\rho}-T^{\rho}_{\rho\mu}T^{\nu\mu}_{\nu}.
\end{equation}
On the other hand, the curvature of the Weitzenb\"{o}ck connection vanishes identically.

~~~~We consider the cosmological model with the following action:
\begin{equation}
S=\int e~ d^{4}x \left[\frac{T}{2 K}+\frac{1}{2}\,\partial_{\mu}\phi
\partial^{\mu}\phi-V(\phi)+\frac{\xi}{2}B(\phi)T\right]+\mathcal{S}_m,
\label{action}
\end{equation}
where $e=\sqrt{-g}$ is the determinant of the tetrad, ~~$K=8\pi G$, ~~$\phi$ is a canonical scalar field, ~~$V(\phi)$ its potential, and $B(\phi)$ its arbitrary nonminimal coupling with the torsion scalar $T$, ~~$\mathcal{S}_m$ is the matter action. In the classical scalar-curvature theory the action has the same form except for the curvature scalar $R$ replacing torsion scalar $T$. 

    For the spatially flat Friedmann-Robertson-Walker tetrad $e^A_\mu = \mathrm{diag}(1, a(t), a(t), a(t))$
\\ (the corresponding metric is $\mathrm{d}s^2=\mathrm{d}t^2-a^2(t)\mathrm{d}l^2$)
the system of field equations is \cite{we}
\begin{equation}
\begin{array}{l}
3H^2=K\left( \frac{{\dot{\phi}}^2}{2}+V(\phi)-3\xi H^2 B(\phi)+\rho\right) ,
\label{system1}
\end{array}
\end{equation}
\begin{equation}
\begin{array}{l}
2\dot{H}=-K( {\dot{\phi}}^2+2\xi H \dot{\phi}B'(\phi)+2\xi\dot{H}B(\phi)+\rho(1+\omega)),
\label{system2}
\end{array}
\end{equation}
\begin{equation}
\begin{array}{l}
\ddot{\phi}+3 H\dot{\phi}+3\xi H^2 B'(\phi)+V'(\phi)=0.
\label{system3}
\end{array}
\end{equation}
Here $a(t)$ is the scale factor, $H(t)\equiv\frac{\dot a}{a}$ is the Hubble parameter, the prime denotes the derivative with respect to $\phi$. We have used that in the chosen tetrad $T=-6H^2$. The matter equation of state is $p=\omega\rho$, where $\omega\in[-1;1]$.

\subsection{The effective potential}
~~~~In the standard scalar-curvature theory of a nonminimal coupling the conformal transformation to Einstein frame is often used. In the Einstein frame the theory is equivalent to GR with a scalar field as a source, so the evolution of the scalar field in an expanding Universe is, as usual, directed to the minimum of the potential, which is not true in the initial nonminimal formulation, called the Jordan frame. This conformal transformation is usually divided into two steps: a redifinition of the potential via $V_{\mathrm{eff}}(\phi)=V(\phi)/U^2(\phi)$, where $U(\phi)=1+K\xi B(\phi)$ in our notations, and a redifinition of the scalar field in order to get the canonical kinetic term. The combination denoted here as $V_{\mathrm{eff}}(\phi)$ is rather interesting by itself. It is a conformal invariant, so it is not changed by any conformal transformation \cite{Kuusk}. Moreover, it can give important information as regards the dynamics of the system without the second step to the Einstein frame (which is usually much more technically complicated than very easily calculated first step). Namely, de Sitter solutions corresponds to minima of the effective potential, their stability is determined by sign of the second derivative of the effective potential in a standard way \cite{wevernov, Kuusk}. 

In the scalar-torsion theory the Einstein frame does not exist \cite{Yang}. However, it is possible to introduce some analog of effective potential with the same properties as in the scalar-curvature theory. Indeed, taking into account (\ref{system1}) we rewrite (\ref{system3}) for $\rho=0$
\begin{equation}
\ddot{\phi}+3 H\dot{\phi}+\frac{K\xi B'(\phi){\dot\phi}^2}{2(1+K\xi B(\phi))}+\frac{K\xi B'(\phi)V(\phi)+V'(\phi)(1+K\xi B(\phi))}{1+K\xi B(\phi)}=0.
\label{ddotphi}
\end{equation}
We see that the effective potential, of the form
\begin{equation}
V_{\mathrm{eff}}(\phi)=V(\phi)(1+K\xi B(\phi)),
\end{equation}
has a derivative with respect to $\phi$
\begin{equation}
\frac{\mathrm{d}V_{\mathrm{eff}}(\phi)}{\mathrm{d}\phi}=V'_{\mathrm{eff}}(\phi)=K\xi B'(\phi) V(\phi)+V'(\phi)(1+K\xi B(\phi)),
\end{equation}
which coincides with numerator of the last term in (\ref{ddotphi}).

The de Sitter solution $H=H_0$, $\phi=\phi_0$, in this model exists for
\begin{equation}
\begin{cases}
3{H_0}^2(1+K\xi B(\phi_0))=KV(\phi_0)
\\3\xi {H_0}^2 B'(\phi_0)=-V'(\phi_0)
\end{cases}
\end{equation}
From this system it is follows that 
\begin{equation}
\begin{array}{l}
\frac{KV(\phi_0)}{1+K\xi B(\phi_0)}=-\frac{V'(\phi_0)}{\xi B'(\phi_0)}~~\Rightarrow\\
\\\Rightarrow K\xi B'(\phi_0) V(\phi_0)+V'(\phi_0)(1+K\xi B(\phi_0))=V'_{eff}(\phi_0)=0
\label{conditiondS}
\end{array}
\end{equation}
Now we add small perturbations to the de Sitter solution: $\phi(t)=\phi_0+\delta\phi$, $\dot\phi(t)=\delta\dot\phi$, 
\\$H(t)=H_0+\delta H$. Substituting these perturbations to (\ref{ddotphi}) we get in the first order
\begin{equation}
\delta\ddot\phi+3 H_0\delta\dot{\phi}+\delta\phi\frac{V''_{\mathrm{eff}}(\phi_0)}{1+K\xi B(\phi_0)}=0.
\end{equation} 
New variables $s=\delta\phi$, $r=\delta\dot\phi$ are introduced and the first-order system of differential equations is written in the form
\begin{equation}
\begin{array}{l}
\dot s=r,\\
\dot r=-3 H_0 r-s\frac{V''_{\mathrm{eff}}(\phi_0)}{1+K\xi B(\phi_0)}.
\label{systems1s2}
\end{array}
\end{equation}
We find eigenvalues for the system (\ref{systems1s2})
\begin{equation}
\begin{array}{l}
\begin{vmatrix}
-\lambda & 1\\
-\frac{V''_{\mathrm{eff}}(\phi_0)}{1+K\xi B(\phi_0)} & -3H_0-\lambda
\end{vmatrix}
=\lambda^2 +3H_0\lambda+\frac{V''_{\mathrm{eff}}(\phi_0)}{1+K\xi B(\phi_0)}=0 \Rightarrow\\
\\\Rightarrow \lambda_{1,2}=-\frac{3}{2}H_0\pm\frac{1}{2}\sqrt{9{H_0}^2-\frac{4 V''_{\mathrm{eff}}(\phi_0)}{1+K\xi B(\phi_0)}}<0 ~~~~\text{for ~~$V''_{\mathrm{eff}}(\phi_0)>0$, ~~$\xi>0$, ~~$B(\phi_0)>0$}.
\end{array}
\end{equation}
So, the de Sitter solution is stable in the minima of the effective potential, as expected. It is worth to note that the second equation in Eq. (\ref{conditiondS}) is just the ``balanced solution'' studied in \cite{Laur}, so from computational point of view the effective potential gives nothing new. However, from heuristic point of view often it is much easier to visualize the locations of the minima of some function instead of doing calculations. For example, for positive power-law potential and coupling functions $B(\phi) \sim \phi^N$ and $V(\phi) \sim \phi^n$ de Sitter solution exists only for negative $n$ with $0< -n < N$. This fact has been established computationally in \cite{we} and becomes now a trivial consequence of the form of $V_{\mathrm{eff}}(\phi)=U(\phi)V(\phi)$.

\subsection{Dimensionless variables}
~~~~We introduce new dimensionless variables
\begin{equation}
\begin{array}{l}
x=\frac{K{\dot{\phi}}^2}{6 H^2(1+K\xi B(\phi))},~~~~
y=\frac{K V(\phi)}{3 H^2(1+K\xi B(\phi))},~~~~
z=\frac{K \rho}{3 H^2(1+K\xi B(\phi))},\\
\\~~~~~~~~~~~~~~~~~~~~~~~~m=\frac{\dot{\phi}}{H \phi},~~~~
A=\frac{\phi B'(\phi)}{1+K\xi B(\phi)}
\label{variables}
\end{array}
\end{equation}
and also dimensionless parameters
\begin{equation}
\begin{array}{l}
b=\frac{\phi B''(\phi)}{B'(\phi)},~~~~
c=\frac{\phi V'(\phi)}{V(\phi)}.
\end{array}
\end{equation}
    Choosing the power-law function $B(\phi)=\phi^N$ and the potential $V(\phi)=V_0 \phi^n$ we get $b=N-1$, $c=n$. 
    
    Note the useful relation between $A$, $x$, $m$, namely,
\begin{equation}
x=\frac{K}{6 N}m^2 A^{\frac{2}{N}}{(N-K\xi A)}^{\frac{N-2}{N}}.
\label{xmAN}
\end{equation}    
    It is useful to introduce auxiliary variables
\begin{equation}
\begin{array}{l}
X=\frac{\ddot{\phi}}{H\dot{\phi}},~~~~
Y=\frac{\dot{H}}{H^2},
\end{array}
\end{equation}
which are expressed through dimensionless variables and parameters from the system (\ref{system1})-(\ref{system3}),
\begin{equation}
\begin{array}{l}
X=-3-\frac{K\xi A m}{2 x}-\frac{c y m}{2 x},\\
\\Y=-3 x-K\xi A m-\frac{3}{2} z(1+\omega),
\label{XYN}
\end{array}
\end{equation}
    From (Eq. \ref{system1}) using (\ref{variables}) we get
\begin{equation}
\begin{array}{l}
1=x+y+z,
\label{z}
\end{array}
\end{equation}
then $z=1-x-y$. 

Taking the derivative of the variables $y$, $m$, $A$ with respect to $\ln (a)$ ($'=\frac{\mathrm{d}}{\mathrm{d}\ln a}$) we obtain the following first-order system of differential equations:
\begin{equation}
\begin{array}{l}
y'=y(cm-2 Y-K\xi A m),\\
\\m'=m(X-Y-m),\\
\\A'=A m(b+1-K\xi A),\
\end{array}
\end{equation}
and finally substituting (\ref{xmAN}), (\ref{XYN}), and (\ref{z}) we get 
\begin{equation}
\begin{array}{l}
y'=y\left[  c m+\frac{K}{b+1} m^2 A^{\frac{2}{b+1}}{(b+1-K\xi A)}^{\frac{b-1}{b+1}}+K\xi m A+\right. \\
~~~~~~\left. +3\left( 1-\frac{K}{6(b+1)} m^2 A^{\frac{2}{b+1}}{(b+1-K\xi A)}^{\frac{b-1}{b+1}}-y\right) (1+\omega)\right]  ,\\
\\m'=-3 m-3(b+1)\xi A^{\frac{b-1}{b+1}}{(b+1-K\xi A)}^{\frac{1-b}{b+1}}-\frac{3 c y (b+1)}{K A^{\frac{2}{b+1}}}{(b+1-K\xi A)}^{\frac{1-b}{b+1}}+\\
~~~~~~~~+\frac{K m^3}{2(b+1)} A^{\frac{2}{b+1}}{(b+1-K\xi A)}^{\frac{b-1}{b+1}}+K\xi A m^2-m^2+\\
~~~~~~~~+\frac{3}{2} m\left( 1-\frac{K m^2}{6(b+1)} A^{\frac{2}{b+1}}{(b+1-K\xi A)}^{\frac{b-1}{b+1}}-y\right) (1+\omega),\\
\\A'=A m(b+1-K\xi A).
\end{array}
\end{equation}

 If $N=2 ~~(b=1)$, then instead of (\ref{xmAN}), and (\ref{XYN}) we have
\begin{equation}
x=\frac{K}{12}m^2 A,
\end{equation}    
\begin{equation}
\begin{array}{l}
X=-3-6\frac{\xi}{m}-6\frac{K c y A}{m},\\
\\Y=-\frac{K}{4}m^2 A-K\xi A m-\frac{3}{2}(1-\frac{K}{12}m^2 A-y)(1+\omega),
\label{XYN2}
\end{array}
\end{equation}
and the first-order system of differential equations
\begin{equation}
\begin{array}{l}
y'=y\Big{(}c m+\frac{K}{2} m^2 A+K\xi m A+3\left( 1-\frac{K}{12} m^2 A-y\right) (1+\omega)\Big{)}  ,\\
\\m'=-3 m-6\xi-\frac{6 c y}{K A}+\frac{K}{4}m^3 A+K\xi A m^2-m^2
+\\
~~~~~~~~+\frac{3}{2} m\left(1-\frac{K m^2}{12} A-y \right)(1+\omega),\\
\\A'=A m(2-K\xi A).
\label{systemyAN2}
\end{array}
\end{equation} 
We consider only the case of $N=2$ in the present paper.

\section{Stationary points and corresponding regimes}
\subsection{Stationary points analysis}
~~~~Solving the system (\ref{systemyAN2}) with vanishing left-hand sides, we find
the following stationary points:
\\
\\
\textbf{1.}~~~~$x=0$, $y=1$, $z=0$, $m=0$, $A=-{\frac{c}{K \xi}}$.
\\We calculate the eigenvalues for the Jacobian matrix associated with the system (\ref{systemyAN2}) to find
\begin{equation}
\begin{array}{l}
\lambda_1 =-3(1+\omega),\\
\lambda_{2,3}=-\frac{3}{2}\pm\frac{1}{2}\sqrt{9-24\xi(c+2)},
\end{array}
\end{equation}
so this point is stable for non-phantom matter.
For this point the quantity $Y=\frac{\dot H}{H^2}=0$ is found using (\ref{XYN2}) and, therefore, $\dot H=0$. Then we find the time dependence of the scale factor,

\begin{equation}
\begin{array}{l}
a(t)=a_0 e^{H_0(t-t_0)}.
\label{dSa}
\end{array}
\end{equation}
As the coordinate of this fixed point is $A=\frac{N\phi^N}{1+K\xi\phi^N}\to-{\frac{n}{K \xi}}\neq0$, then the scalar field approaches a constant 
\begin{equation}
\begin{array}{l}
\phi=\phi_0.
\label{dSphi}
\end{array}
\end{equation}
This is a de Sitter solution. The constants $\phi_0$ and $H_0$ are found with the substitution of this solution to the initial system of Eqs. (\ref{system1})-(\ref{system3}): ~~$\phi_0=\pm\sqrt{-\frac{n}{K\xi(2+n)}}$, ~~$H_0=\pm\sqrt{-\frac{nV_0{\phi_0}^{n-2}}{6\xi}}$. Clearly, this regime exists only for $0>n>-2$. Since it is stable it can be used for describing late-time acceleration of our Universe. Definitely, the unnaturally low value of $H_0$ (in natural units) should transform to very low value of some of the parameters of the theory. Note, however, that in the theory under investigation we have more possibilities to get a very low $H_0$ --- this may happen due to either very low $V_0$, a very low $\xi$ or a potential power index $n$ being very close to $-2$. For the last possibility, if we denote $n+2=\epsilon << 1$, we get $H_0 \sim \epsilon\sqrt{V_0 \xi}$, so small corrections to the $\phi^{-2}$ 
potential would do the job.
\\
\\
\textbf{2.}~~~~$x=1$, $y=0$, $z=0$, $m=\sqrt{6\xi}$, $A=\frac{2}{K\xi}$.
\\The corresponding eigenvalues are
\begin{equation}
\begin{array}{l}
\lambda_1=-2\sqrt{6\xi},\\
             \lambda_2=3(1-\omega)+2\sqrt{6\xi},\\
                    \lambda_3=(c+2)\sqrt{6\xi}+6.
\end{array}
\end{equation}                    
    Using (\ref{XYN2}) the quantity $Y$ at this stationary point is obtained to be $Y_{\mathrm{stat}}=-3-2\sqrt{6\xi}$ and now we can restore the time dependence of the Hubble parameter, $H(t)=-\frac{1}{Y_{\mathrm{stat}}(t-t_0)}$, and the scale factor     
\begin{equation}
\begin{array}{l}
a(t)=a_0{|t-t_0|}^{-\frac{1}{Y_{\mathrm{stat}}}}=a_0{|t-t_0|}^{\frac{1}{3+2\sqrt{6\xi}}}.
\end{array}
\end{equation}
The corresponding behavior of the scalar field is obtained applying the definition of the variable $m=\frac{\dot\phi}{H\phi}$, then ~~$\frac{\dot\phi}{\phi}=m_{\mathrm{stat}}\frac{\dot a}{a}$, where $m_{\mathrm{stat}}$ -- the coordinate of a stationary point,
\begin{equation}
\begin{array}{l}
\phi(t)=\phi_0{|t-t_0|}^{-\frac{m_{\mathrm{stat}}}{Y_{\mathrm{stat}}}}=\phi_0{|t-t_0|}^{\frac{\sqrt{6\xi}}{3+2\sqrt{6\xi}}}.
\end{array}
\end{equation}
\\
\\\textbf{3.}~~~~$x=1$, $y=0$, $z=0$, $m=-\sqrt{6\xi}$, $A=\frac{2}{K\xi}$.\\
The eigenvalues for this fixed point are
\begin{equation}
\begin{array}{l}
\lambda_1 =2\sqrt{6\xi},\\
\lambda_2=3(1-\omega)-2\sqrt{6\xi},\\
\lambda_3=6-\sqrt{6\xi}(c+2).
\end{array}
\end{equation}

    We get asymptotic behavior $a(t)$ and $\phi(t)$ calculating the quantity $Y_{\mathrm{stat}}=-3+2\sqrt{6\xi}$ and using $m=m_{\mathrm{stat}}=-\sqrt{6\xi}$ at this point,
\begin{equation}
\begin{array}{l}
a(t)=a_0{|t-t_0|}^{\frac{1}{3-2\sqrt{6\xi}}},\\
\\\phi(t)=\phi_0{|t-t_0|}^{\frac{\sqrt{6\xi}}{2\sqrt{6\xi}-3}}.
\label{aphi3}
\end{array}
\end{equation}
For $\xi=\frac{3}{8}$ instead of this solution (\ref{aphi3}) we have the exponential one (because in this case $Y=0$, ~~$H=H_0$, ~~$\frac{\dot\phi}{\phi}=m_{\mathrm{stat}}H_0$, ~~$m_{\mathrm{stat}}=-\frac{3}{2}$)
\begin{equation}
\begin{array}{l}
a(t)=a_0e^{H_0(t-t_0)},\\
\\\phi(t)=\phi_0 e^{-\frac{3}{2} H_0(t-t_0)}.
\label{exp3aphi}
\end{array}
\end{equation}

The points $2$ and $3$ represent a situation when kinetic term of the scalar field dominates.
\\
\\\textbf{4.}~~~~$x=\frac{8\xi}{3(\omega-1)^2}$, $y=0$, $z=\frac{3+3\omega^2-6\omega-8\xi}{3(\omega-1)^2}$, $m=\frac{4\xi}{\omega-1}$, $A=\frac{2}{K\xi}$.
\\We find eigenvalues for this point,
\begin{equation}
\begin{array}{l}
\lambda_1=\frac{3\omega^2-3+4 c\xi}{\omega-1}\\
\lambda_2=\frac{3{(\omega-1)}^2-8\xi}{2(\omega-1)}\\
\lambda_3=\frac{8\xi}{1-\omega}.
\end{array}
\end{equation}
As the quantity $Y$ at this stationary point is $Y_{\mathrm{stat}}=\frac{3-3\omega^2-8\xi}{2(\omega-1)}$, we have
\begin{equation}
\begin{array}{l}
a(t)=a_0{|t-t_0|}^{\frac{2(\omega-1)}{3\omega^2-3+8\xi}},\\
\\\phi(t)=\phi_0{|t-t_0|}^{\frac{8\xi}{3\omega^2-3+8\xi}}.
\label{aphi4}
\end{array}
\end{equation}
The corresponding behavior $\rho(t)$ is found using $\dot \rho+3H\rho(1
+\omega)=0$, then $\frac{\dot \rho}{\rho}=-3(1+\omega)\frac{\dot a}{a}$ and 
\begin{equation}
\begin{array}{l}
\rho(t)=\rho_0{|t-t_0|}^{\frac{3(1+\omega)}{Y_{\mathrm{stat}}}}=\rho_0{|t-t_0|}^{\frac{6(1-\omega^2)}{3\omega^2-3+8\xi}}.
\label{rho4}
\end{array}
\end{equation}
This is a kinetic tracker regime when the potential term is negligible, and the ratio between the kinetic energy of the scalar field and the matter
density remains constant. When $\xi=\frac{3}{8}(1-\omega^2)$, the solution (\ref{aphi4}), (\ref{rho4}) does not exist and the exponential one appears ($m_{\mathrm{stat}}=-\frac{3}{2}(1+\omega)$),
\begin{equation}
\begin{array}{l}
a(t)=a_0 e^{H_0(t-t_0)},\\
\\\phi(t)=\phi_0 e^{-\frac{3}{2}(1+\omega)H_0(t-t_0)},\\
\\\rho(t)=\rho_0 e^{-3(1+\omega)H_0(t-t_0)},
\label{exp4aphirho}
\end{array}
\end{equation}
where ${H_0}^2=-\frac{4\rho_0}{9{\phi_0}^2\omega(1+\omega)}$ is found on substituting the exponential solution (\ref{exp4aphirho}) into the initial system of equations (\ref{system1})-(\ref{system3}).
\\
\\\textbf{5.}~~~~$x=\frac{\xi{(c+2)}^2}{6}$, $y=1-\frac{\xi{(c+2)}^2}{6}$, $z=0$, $m=-\xi(c+2)$, $A=\frac{2}{K\xi}$.
\\
The eigenvalues are calculated,
\begin{equation}
\begin{array}{l}
\lambda_1 =2\xi(c+2),\\
\lambda_2=-3+\frac{\xi{(c+2)}^2}{2},\\
\lambda_3=-3(1+\omega)+c\xi(c+2).
\end{array}
\end{equation}

    Similar to the previous points, using ~~$Y_{\mathrm{stat}}=\frac{1}{2}\xi(4-c^2)$ and ~~$m_{\mathrm{stat}}=-\xi(c+2)$ we find $a(t)$, $\phi(t)$ ~~($c=n$),
\begin{equation}
\begin{array}{l}
a(t)=a_0{|t-t_0|}^{\frac{2}{\xi(n^2-4)}},\\
\\\phi(t)=\phi_0{|t-t_0|}^{{\frac{2}{2-n}}}.
\end{array}
\end{equation}
\\
\\\textbf{6.}~~~~$x=\frac{3{(1+\omega)}^2}{2\xi c^2}$, $y=\frac{3-3\omega^2-4 c\xi)}{2\xi c^2}$, $z=\frac{c^2\xi-3-3\omega+2 c\xi}{\xi c^2}$, $m=\frac{-3(1+\omega)}{c}$, $A=\frac{2}{K\xi}$.
\\We obtain the eigenvalues for this point,
\begin{equation}
\begin{array}{l}
\lambda_1=\frac{6(1+\omega)}{c},\\
   \\\lambda_{2,3}=\frac{1}{4\xi c}\left( 3\xi(2-c+\omega(c+2))\pm\sqrt{-3\xi(72\omega^3+f_1(\xi,c)\omega^2+f_2(\xi,c)\omega+f_3(\xi,c))}\right) .
\label{lambda6}
\end{array}
\end{equation}
where $f_1(\xi,c)=-27 c^2\xi-12\xi-60 c\xi+72$,
\\\text{~~~~~~~~}$f_2(\xi,c)=-24\xi-72+96 c\xi+6 c^2\xi$, and
\\\text{~~~~~~~~}$f_3(\xi,c)=21 c^2\xi-72+156 c\xi-32\xi^2 c^3-12\xi-64 c^2\xi^2$.

The quantities $Y_{\mathrm{stat}}=\frac{3}{2 c}(1+w)(2-c))$ and $m_{\mathrm{stat}}=\frac{-3(1+\omega)}{c}$ give us the behavior $a(t)$, $\phi(t)$, and $\rho(t)$ ~~($c=n$)
\begin{equation}
\begin{array}{l}
a(t)=a_0{|t-t_0|}^{\frac{2 n}{3(1+\omega)(n-2)}},\\
\\\phi(t)=\phi_0{|t-t_0|}^{\frac{2}{2-n}},\\
\\\rho(t)=\rho_0{|t-t_0|}^{\frac{2 n}{2-n}},
\end{array}
\end{equation}
This is a tracker when the potential and kinetic terms of the scalar field as well as the density of the matter are constant with respect to each other.

\subsection{Conditions of existence and stability of obtained solutions}
~~~~~With the substitution of the power-law solution ~~$a(t)=a_0{(t-t_0)}^\alpha$, ~~$\phi(t)=\phi_0{(t-t_0)}^{\alpha\beta}$, 
\\and $\rho(t)=\rho_0{(t-t_0)}^{-3(1+\omega)\alpha}$ to the initial system of equations (\ref{system1}), (\ref{system2}), (\ref{system3}), where $B(\phi)=\phi^2$, $V(\phi)=V_0\phi^n$, we find the conditions of the existence and stability of the six solutions obtained from the previous subsection. We consider only the cases with $n<0$, $\xi>0$, and only positive values of $\phi$. It is worth to recall that in general a fixed point does not necessary correspond to some solution of the initial system. First of all, all regimes found above (except the de Sitter solution, which is an exact, and not only an asymptotic solution) exist in the $\phi \to \infty$ limit where we can neglect the Einstein term in comparison with the term originating from a nonminimal coupling in the denominator of the expansion in normalized variables we use. When the field is small (which is realized near a cosmological singularity for decreasing power-law potentials) the Universe expands according to GR, since the correction terms are less important. Such regimes for quintessence potentials are well known and are not included in our analysis here. As for the large $\phi$ regimes, studied in the present paper, we require that the omitted terms in the equations of motion be negligible for large $\phi$. Otherwise the regime cannot be realized as an asymptotic solution. For example, suppose we have a vacuum asymptotic solution. If the influence of matter grows with growing $\phi$, this means that the asymptotic solution corresponding to this particular expansion dynamics is absent if we add any amount of matter. This does not mean that such a regime has no physical meaning at all --- the Universe may follow it as transient one, if the amount of matter is small enough. However, such a situation does not belong to the asymptotic regimes, and we will not list it in the following.

The analysis described above (with stability results got from the corresponding eigenvalues) leads to the following results for vacuum solutions summarized in Table 1 (points $4$ and $6$, being non-vacuum, are not included).

We can see from this table that in the vacuum case there are three possible future stable regimes:
\begin{itemize}
\item For $n>-2$ a de Sitter solution exists and is stable.
\item For $n<-2$ and $\xi < 6/{(n+2)}^2$ the regime of point $5$ is stable.
\item For $n<-2$ and $\xi>6/{(n+2)}^2$ the kinetic dominating regime of point $2$ is stable.
\end{itemize}

When the matter is taken into account, the situation is described in the Table 2 (note that the de Sitter solution in the presence of matter is an asymptotic solution).

\newpage
\begin{center}
Table 1: \textbf{Conditions of existence and stability of solutions in vacuum}
\begin{tabular}{|c|c|c|c|}
\hline
{} & \multicolumn{2}{c|}{} & {} \\
\textbf{\textnumero} & \multicolumn{2}{c|}{\textbf{Conditions}} & \textbf{Type}\\
{\textbf{of point}} & \multicolumn{2}{c|}{\textbf{of existence}} & ~~\textbf{of stability}~~\\
\hline
{} & \multicolumn{2}{c|}{} & {}\\
{} & \multicolumn{2}{c|}{} & $0<\xi\leqslant\frac{3}{8(n+2)}$ \\
\textbf{1.} & \multicolumn{2}{c|}{$-2<n<0$} & \textbf{Stable} node\\
\cline{4-4}
{} & \multicolumn{2}{c|}{} & {}\\
{} & \multicolumn{2}{c|}{} & $\xi>\frac{3}{8(n+2)}$\\
{} & \multicolumn{2}{c|}{} & \textbf{Stable} focus\\
\hline
{} & \multicolumn{2}{c|}{} & {}\\
\textbf{2.} &  \multicolumn{2}{c|} {$n<-2$, ~~$\xi>\frac{6}{{(n+2)}^2}$, ~~$t\to\infty$} & \textbf{Stable} node\\
{} & \multicolumn{2}{c|}{} & {}\\
\hline 
{} & {} & {} & {}\\
{} & {} & \textbf{1).} $0<\xi<\frac{3}{8}$, ~~$t\to t_0$ & {}\\
{} & $-2<n<0$, & ~~~~~~\textbf{2).} $\frac{3}{8}<\xi<\frac{6}{{(n+2)}^2}$, ~~$t\to\infty$ & {}\\
{} & {} & {} & {}\\
\cline{2-3}
\textbf{3.} & {} & {} & {}\\
{} & {}  & \textbf{1).} $0<\xi<\frac{3}{8}$, ~~$t\to t_0$ & Unstable node\\
{} & $n\leqslant-2$, &\textbf{2).} $\xi>\frac{3}{8}$, ~~$t\to\infty$ & {}\\
{} & {} & {} & {}\\
\cline{2-3}
{} & \multicolumn{2}{c|}{} & {}\\
{} & \multicolumn{2}{c|}{$n<0$, ~~$\xi=\frac{3}{8}$, ~~$t\to-\infty$} & {}\\
{} & \multicolumn{2}{c|}{} & {}\\
\hline
{} & \multicolumn{2}{c|}{} & {}\\
{} & \multicolumn{2}{c|}{$-2<n<0$, ~~$0<\xi<\frac{6}{{(n+2)}^2}$, ~~$t\to\infty$} & Saddle\\
\textbf{5.} & \multicolumn{2}{c|}{} & {}\\
\cline{2-4}
{} & \multicolumn{2}{c|}{} & {}\\
{} & \multicolumn{2}{c|}{$n<-2$, ~~$0<\xi<\frac{6}{{(n+2)}^2}$, ~~$t\to\infty$} & \textbf{Stable} node\\
{} & \multicolumn{2}{c|}{} & {}\\
\hline
\end{tabular}
\end{center}

\newpage
\begin{center}
Table 2: \textbf{Conditions of existence and stability of solutions  for $\rho\neq0$ }

\begin{tabular}{|c|c|c|c|}
\hline
{} & \multicolumn{2}{c|}{} & {} \\
\textbf{\textnumero} & \multicolumn{2}{c|}{\textbf{Conditions}} & \textbf{Type}\\
{\textbf{of point}} & \multicolumn{2}{c|}{\textbf{of existence}} & ~~\textbf{of stability}~~\\
\hline
{} & \multicolumn{2}{c|}{} & {}\\
{} & \multicolumn{2}{c|}{} & $0<\xi\leqslant\frac{3}{8(n+2)}$ \\
\textbf{1.} & \multicolumn{2}{c|}{$-2<n<0$, ~~$\omega\in(-1;1]$} & \textbf{Stable} node\\
\cline{4-4}
{} & \multicolumn{2}{c|}{} & {}\\
{} & \multicolumn{2}{c|}{} & $\xi>\frac{3}{8(n+2)}$\\
{} & \multicolumn{2}{c|}{} & \textbf{Stable} focus\\
\hline 
{} & {} & {} & {}\\
{} & {} & \textbf{1).} $\omega\in[-1;0)$, ~~$0<\xi<\frac{3{(1-\omega)}^2}{8}$, ~~$t\to t_0$ & {}\\
\textbf{3.} & $n<0$, &\textbf{2).} $\omega\in[0;1)$, ~~$\frac{3}{8}<\xi<\frac{6}{{(n+2)}^2}$, ~~$t\to\infty$ &  {}\\
{} & {} & {} & Unstable node\\
\cline{2-3}
{} & \multicolumn{2}{c|}{} & {}\\
{} & \multicolumn{2}{c|}{$n<0$, ~~$\xi=\frac{3}{8}$, ~~$\omega\in[-1;0)$, ~~$t\to-\infty$} & {}\\
{} & \multicolumn{2}{c|}{} & {}\\
\hline
{} & \multicolumn{2}{c|}{} & {}\\
{} &  \multicolumn{2}{c|}{$n<0$, ~~$0<\xi<\frac{3(1-\omega^2)}{8}$, ~~$\omega\in(-1;1)$, ~~$t\to t_0$} &  Unstable node\\
\textbf{4.} & \multicolumn{2}{c|}{} & {}\\
\cline{2-4}
{} & \multicolumn{2}{c|}{} & {}\\
{} & \multicolumn{2}{c|}{$n<0$, ~~$\xi=\frac{3}{8}(1-\omega^2)$, ~~$\omega\in(-1;0)$, ~~$t\to-\infty$} & Saddle\\
{} & \multicolumn{2}{c|}{} & {}\\
\hline
{} & \multicolumn{2}{c|}{} & {}\\
\textbf{5.} & \multicolumn{2}{c|}{$n<-2$, ~~$0<\xi<\frac{3(1+\omega)}{n(n+2)}$, ~~$\omega\in(-1;1]$, ~~$t\to\infty$} & \textbf{Stable} node\\
{} & \multicolumn{2}{c|}{} & {}\\
\hline
{} & \multicolumn{2}{c|}{} & {}\\
{} & \multicolumn{2}{c|}{} & $\frac{3(1+\omega)}{n(n+2)}<\xi\leqslant\xi_0$\\
\textbf{6.} & \multicolumn{2}{c|}{$n<-2$, ~~$\xi>\frac{3(1+\omega)}{n(n+2)}$, ~~$\omega\in(-1;1]$, ~~$t\to\infty$} & \textbf{Stable} node \\
\cline{4-4}
{} & \multicolumn{2}{c|}{} & {}\\
{} & \multicolumn{2}{c|}{} & $\xi>\xi_0$\\
{} & \multicolumn{2}{c|}{} & \textbf{Stable} focus \\
\hline
\end{tabular}
\end{center}

In the Table 2 ~~$\xi_0=-\frac{3}{64n^2(n+2)}(f_4(n,\omega)+\sqrt{f_5(n,\omega)f_6(n,\omega)})$,
\\where $f_4(n,\omega)=-52n-7n^2+4+\omega^2(9n^2+20n+4)+2\omega (4-16n-n^2)$
\\\text{~~~~~~~~}$f_5(n,\omega)=\omega^2(4+36n+81n^2)+\omega(126n^2-64n+8)+4-100n+49n^2$,
\\\text{~~~~~~~~}$f_6(n,\omega)={(2-n+\omega(n+2))}^2$.

For $\xi=\xi_0$ the square root equals zero in the eigenvalues $\lambda_{2,3}$ of point 6 (see \ref{lambda6}).

We see that point $2$ (a kinetic dominating regime) is absent --- this situation is described in the beginning of this section as an illustration of what we call the existing asymptotic regimes. In this particular case matter with $w<1$ would inevitably destroy a vacuum regime of the point $2$ in late time. The future stable regimes are:

\begin{itemize}
\item For $n>-2$, as for the vacuum case, the future stable point is de Sitter solution.
\item For $n<-2$ and $\xi<3(1+\omega)/(n^2+2n)$ the regime of point $5$ is stable, so the scalar field dominates in late time.
\item For $n<-2$ and $\xi>3(1+\omega)/(n^2+2n)$ the tracker (point $6$) is stable to the future.
\end{itemize}

\begin{figure}[hbtp] 
~~~~~~~~~~~~\includegraphics [scale=0.68] {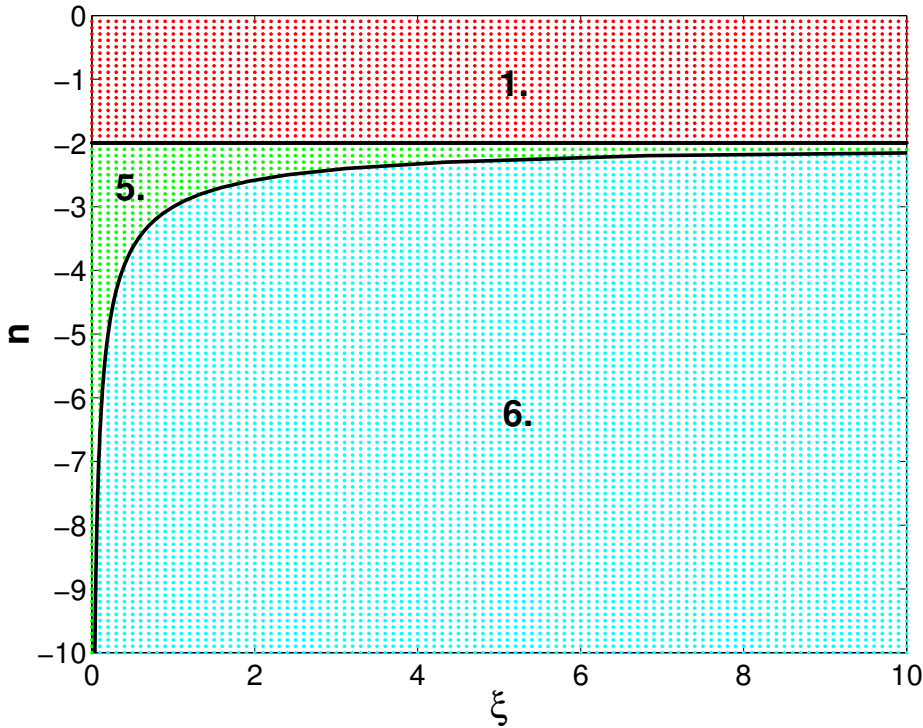}
\caption{Regions of a stability of solutions corresponding to the points 1 (red), 5 (green), 6 (cyan) for the case  
$\rho\neq0$ with $\omega=0$.}
\label{Fig1}
\end{figure}

An example of configurations of stability regions for $\omega=0$ is shown in Fig. \ref{Fig1}.  

\section{Conclusions}
~~~~In the present paper we considered the evolution of the Universe in a teleparallel version of a nonminimally coupled scalar field theory, focusing mostly on quintessence scalar field potentials. In our previous paper \cite{we} we already pointed out the radical differences in the dynamics of a ``classical'' and teleparallel cases of this theory. Here we provided a qualitative explanation of this difference using the notion of an effective scalar field potential. It appears as a first step from a Jordan to an Einstein frame (to be followed by redefinition of the scalar field in order to transform the kinetic term to the canonical form) in the standard theory, though it already gives an intuitively clear picture of possible dynamical regimes without this second step. It is interesting that in the teleparallel version, where the Einstein frame does not exist at all \cite{Yang}, the effective potential nevertheless can be introduced in the same manner. The difference in the forms of this effective potential (being $V(\phi)/U^2(\phi)$ for the classical case and $V(\phi)U(\phi)$ for the teleparallel case, where $V(\phi)$ is the scalar field potential and $U(\phi)$ is the coupling function which includes the Einstein-Hilbert term) results in very different dynamics for the same scalar field potentials. 

For the standard scalar-curvature coupling, the presence of $U(\phi)$ [which is usually taken in the form $U(\phi)=1+8\pi G\xi \phi^2$] in the denominator leads to run-away solution for growing $V(\phi)$, corresponding to $V_{\mathrm{eff}}(\phi)$ decreasing for large $\phi$. This may happen if $U^2(\phi)$ grows more rapidly than $V$ with $\phi$ \cite{wevernov}. In the boundary case of $U^2(\phi) \sim V(\phi)$ for large $\phi$ the effective potential is asymptotically flat, which is the reason for the existence of viable Higgs inflation models \cite{Bezrukov}. 

None of these features can exist for the effective potential in the form of $U(\phi)V(\phi)$ as in scalar-torsion theory. For the usual quadratic coupling (we study only this form of the coupling in the present paper) $V_{\mathrm{eff}}(\phi)$ is growing for an arbitrarly growing $V(\phi)$, so the dynamics inevitably pushes the scalar field toward zero, as we have seen in our previous paper. On the contrary, for the decreasing potentials studied in the present paper the effective potential can have a minimum, resulting in de Sitter solution. For a quadratic coupling this happens if $V(\phi)$ scales less steep than $\phi^{-2}$. This condition for a de Sitter solution to exist has been obtained earlier in \cite{we} and in a more general context in \cite{Laur}, and we now have a very easy way to explain it. 

Steeper potentials result in unlimited growing of the scalar field, similar to the quintessence scenario in GR. We have identified a tracker solution which is always an attractor in its range of existence. It is shown that when the coupling constant $\xi$ is low enough for the tracker to exist (see the corresponding condition in Table 2), a vacuum regime is stable, so the scalar field dominates at late time. We also list several unstable asymptotic regimes. A full description of the scalar field  dynamics requires matching solutions studied in the present paper with GR solutions valid for small $\phi$. We leave this for future work. 

\section*{Acknowledgements}
~~~~The work was supported by RSF Grant \textnumero 16-12-10401 and by the Russian Government
Program of Competitive Growth of Kazan Federal University. Authors are grateful to Emmanuel Saridakis for discussions.

\end{document}